\documentclass{llncs}

\usepackage{epsfig}
\usepackage{epstopdf}

\usepackage[font=small,skip=0pt]{caption}
\begin{document}
\title{Protein Folding in the Hexagonal Prism Lattice with Diagonals}


\subtitle{}


\author{Dipan Lal Shaw \and M. Sohel Rahman \and A. S. M. Shohidull Islam \and Shuvasish Karmaker}


\institute{A$\ell$EDA Group, CSE, BUET, Bangladesh. \and Department of CSE, BUET, Dhaka, Bangladesh}



\maketitle

\begin{abstract}
Predicting protein secondary structure using lattice model is one of the most studied computational problem in bioinformatics. Here secondary structure or three dimensional structure of protein is predicted from its amino acid sequence. Secondary structure refers to local sub-structures of protein. Mostly founded secondary structures are alpha helix and beta sheets. Since, it is a problem of great potential complexity many simplified energy model have been proposed in literature on basis of interaction of amino acid residue in protein. Here we use well researched Hydrophobic-Polar (HP) energy model. In this paper, we proposed hexagonal prism lattice with diagonal that can overcome the problems of other lattice structure, e.g., parity problem.  We give two approximation algorithm for protein folding on this lattice. Our first algorithm leads us to similar structure of helix structure which is commonly found in protein structure. This motivated us to find next algorithm which improves the algorithm ratio of $\frac{9}{7}$.
\end{abstract}

\section{Introduction}
Protein structure prediction is one of the most studied computational problems in bioinformatics. By using simplified and abstract models, many approximate solutions for this problem have been given in the literature. There exist a variety of models attempting to simplify the problem by abstracting only the ``essential physical properties" of real proteins. A lattice model for folding amino acids is represented by connected beads in two dimensional lattices or three dimensional cubic lattices and considers a simplified energy function. \\
We can categorize the lattice structure models into two different classes: Simplified Lattice Models (e.g. \cite{14}) and Realistic Lattice Models \cite{26}. One of the widely used simplified lattice model is the HP model which was first introduced by Dill \cite{14}. In HP model, there are only two types of beads: H represents a hydrophobic or non-polar bead and P represents a polar or hydrophilic one. The main force in the folding process is the hydrophobic-hydrophobic force, i.e., H-H contacts. For optimal embedding, our main goal in this model is to maximize the H-H contacts.\\
The protein folding problem in HP model is NP-hard \cite{15}. Hart and Istrail gave the first 4-approximation algorithm for the problem on the 2D square lattice \cite{17}. Later on, Newman \cite{19} improved the approximation ratio to 3 considering the conformation as a folded loop. A $\frac{8}{3}$-approximation algorithm for the problem on the 3D square lattice was given by Hart and Istrail \cite{17}. In \cite{24}, the authors introduced square lattice with diagonals and presented algorithms that give an approximation ratio of $\frac{26}{15}$ for the two-dimensional and $\frac{8}{5}$ for the three-dimensional lattice. Later, Newman and Ruhl improved this based on different geometric ideas; they achieved an improved approximation ratio of 0.37501 \cite{20}. To remove the parity problem of the square and cubic lattices Agarwala et al. first proposed the triangular lattice in \cite{23}. There, they gave a $\frac{11}{6}$ approximation algorithm. For a more generalized version, namely, the 3D FCC lattice, Agarwala et al. \cite{23} gave an approximation algorithm having an approximation ratio of $\frac{5}{3}$. To alleviate the problem of sharp turns, Jiang and Zhu introduced the hexagonal lattice model and gave an approximation algorithm with approximation ratio 6 \cite{25}. A linear time approximation algorithm for protein folding in the HP side chain model on the extended cubic lattice having an approximation ratio of 0.84 was presented by Heun \cite{Heun99}. \\
A number of heuristic and meta-heuristic techniques have also been applied to tackle the protein folding problem in the literature. A genetic algorithm for the protein folding problem in the HP model in 2D square lattice was proposed in \cite{Unger}. In~\cite{Hoque06,Hoque07}, a hybrid genetic algorithm was presented for the HP model in 2D triangular lattice and 3D FCC lattice. The authors in~\cite{Lesh} first proposed the $pull~ move~ set$ for the rectangular lattices, which was used in the HP model under a variety of local search methods. They also showed the completeness and reversibility of the pull move set for the rectangular grid lattices. In~\cite{Dayem}, the authors extended the idea of the $pull~ move~ set$ in the local search approach for finding an optimal embedding in 2D triangular grid and the FCC lattice in 3D.\\
In this paper, we introduce the hexagonal prism lattices with diagonals for protein folding. Our prior work on hexagonal lattice model with diagonals gives an approximation ratio of $\frac{5}{3}$ for primary protein structure \cite{BMC}. The motivation for introducing hexagonal prism lattice comes from the secondary structure of a protein as follows. The secondary structure of a protein suggests that, in real protein folding, sharp turn does not occur frequently. Hexagonal model alleviates this sharp turn problem \cite{25}. On the other hand, in the cubic lattice HP model there is a serious shortcoming, namely, the $parity~problem$ as follows. Due to a grid structure in a cubic lattice, contact can be established between two hydrophobic atoms only if they both are either on even positions or on odd positions of the sequence. To address this $parity~problem$, we propose idea of this new lattice model, i.e., hexagonal prism lattice model with diagonals. In this model contacts may exist through diagonals (see Fig. \ref{fig1}). Notably, these issues have also been partially alleviated in the cubic lattice with diagonals and triangular lattice. To this end, our new model opens a new avenue for further research for this long standing problem. We present two novel approximation algorithms for long structure protein folding on this lattice. Our first algorithm provide $2$ approximation ratio for $k > 13$ where $k$ is the number of sequences of H's in the HP string. Our next algorithm improves the approximation ratio to $\frac{9}{7}$ for $k > 132$ where $k$ is the number of sequences of H's in the HP string. This algorithm is based on a strategy of partitioning the entire protein sequence into two pieces. Since now Alireza Hadj Khodabakhshi et al. used hexagonal prism lattice more successfully for finding inverse protein folding, which is due to \cite{Alireza}. 

The rest of the paper is organized as follows. In Section `\nameref{Preliminaries}', we introduce the hexagonal prism lattice with diagonals and define some related notions. Section `\nameref{Our Approach}' describes our algorithms and relevant results. We briefly conclude in Section `\nameref{Conclusion}. 

\section{Preliminaries}
\begin{figure}
\centering
\begin{minipage}{.56\textwidth}
\centering
{\epsfig{file=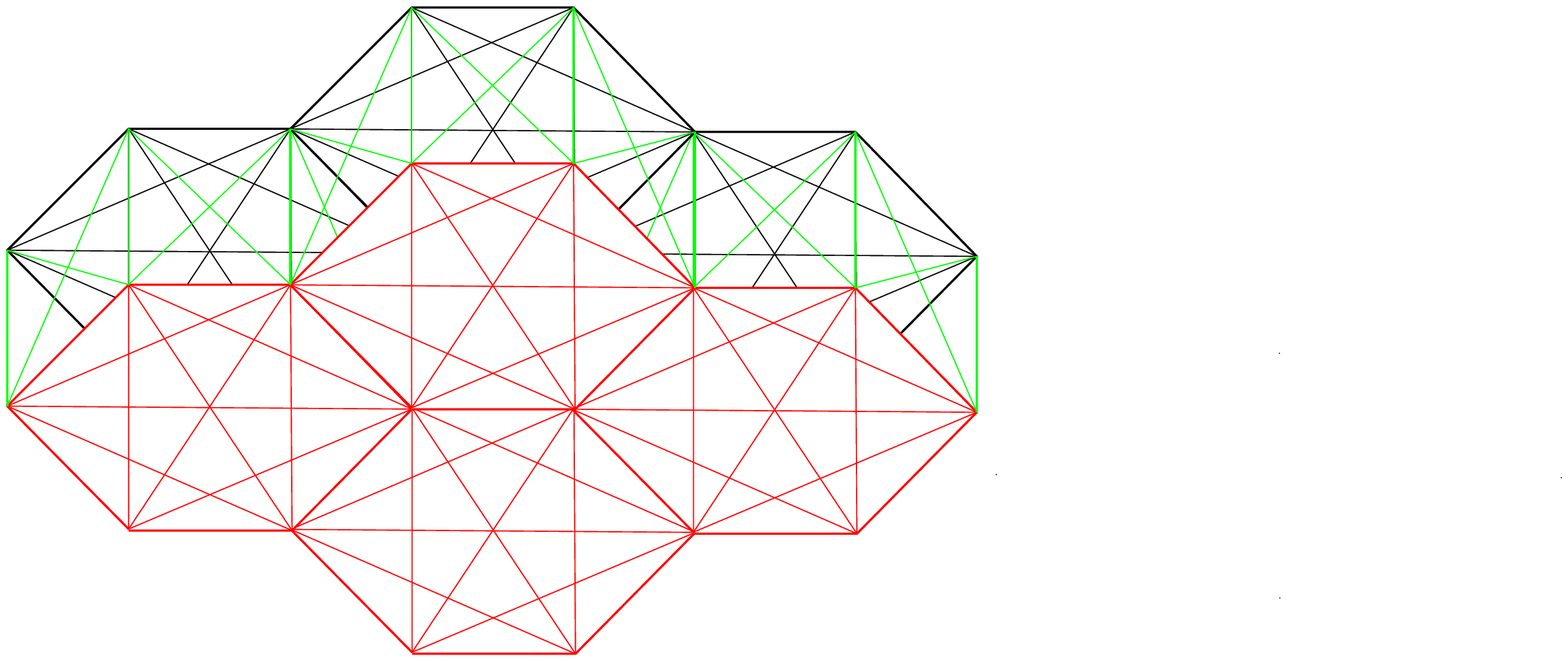,scale=.3}}
\caption{{\normalsize A hexagonal prism lattice with diagonals. Different layers are indicated using black and red color. Connecting edges between layers are indicated using green color.}}\label{fig1}
\end{minipage}
\hfill
\begin{minipage}{.4\textwidth}
\centering
{\epsfig{file=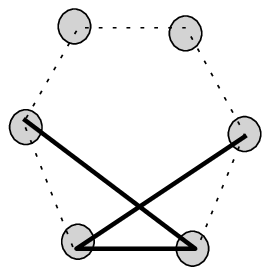,scale=.8}}
\caption{{\normalsize Crossing between binding edges; this situation is forbidden in a valid conformation.}}\label{fig2}
\end{minipage}
\end{figure}

In this section, we present the required notions and notations to describe the hexagonal prism lattice model with diagonals.
\begin{definition}
 The three-dimensional hexagonal prism lattice with diagonals is an infinite graph $G=(V,E)$ in the Euclidian Space with vertex set $V = {R}^3$ and edge set $E = \lbrace (x,x^{\prime}) | x,x^{\prime} \in {R}^3$  $, |x - x^{\prime}| \leq 2 \rbrace $, where $|.| $  denotes the Euclidean norm. The hexagonal prism lattice is composed by stacking multiple two-dimensional hexagonal lattices with diagonals on top of each other. On a hexagonal prism lattice with diagonals each two-dimensional hexagonal lattice with diagonals is called a layer. The edges connecting the two layers are called layer edges. An edge $e \equiv (x,x^{\prime}) \in E $ is a non-diagonal or non-diagonal layer edge iff $|x-x^{\prime}|=1$; otherwise it is a diagonal edge or diagonal-layer edge.
\end{definition}
We use the well known notion of neighbourhood or adjacency of graph theory: two vertices are adjacent/neighbour to each other if they are connected through an edge. In this connection, the difference between the usual hexagonal prism model and our propose model lies in the fact that a vertex in the former has 5 neighbours, whereas in the latter it has additional 15 neighbours, i.e., a total of 20 neighbours (see Fig. \ref{fig1}).

Although the lattice is defined as an infinite graph, we will be concerned with only a finite sub-graph of it for each conformation of a protein. The input to the protein folding problem is a finite string $p$ over the alphabet $\lbrace P,~H \rbrace$ where $p ~ =\lbrace P \rbrace^*b_1 \lbrace P \rbrace^+b_2\lbrace P \rbrace^+ ... \lbrace P \rbrace^+b_k\lbrace P \rbrace^* $. Here $b_i \in \lbrace{H} \rbrace^+ ~ for ~ 1 \leq i \leq k$ and let $n = \sum_{i=1}^{k} |b_i| $. Here, H denotes non-polar and P denotes polar amino acids respectively. Often, in what follows, the input string in our problem will be refer to as an HP string. An H-run in an HP string denotes the consecutive H's and a P-run denotes consecutive P's. So, the total number of H-runs is $k$ and total number of H is $n$. An H-run of even (odd) length is said to be an even H-run (odd H-run). We will now define the valid embeddings and conformation of a protein into this lattice. An embedding is a \textbf{self-avoiding walk} inside the grid.

\begin{definition}
 Let $p$ = $p_1$ $\ldots $ $p_t$ be an HP string of length $t$ and let  $G=(V,~E)$ be a lattice. An embedding of $p$ into $G$ is a mapping function $f$: $\lbrace 1, \ldots,t  \rbrace $ $ \rightarrow V$  from the positions of the string to the vertices of the lattice. It assigns adjacent positions in $p$ to adjacent vertices in $G$, $ (f(i),f(i + 1))$ $ \in E$ for all $1 \leq i \leq t-1 $. The edges $(f(i),f(i + 1)) \in E$ for $ 1 \leq i \leq t-1 $ are called \textbf{binding edges}. An embedding of $p$ into $G$ is called a conformation, if no two binding edges cross each other (see Fig. \ref{fig2}). 
\end{definition}

\begin{figure}
\centering
\begin{minipage}{.45\textwidth}
\centering
{\epsfig{file=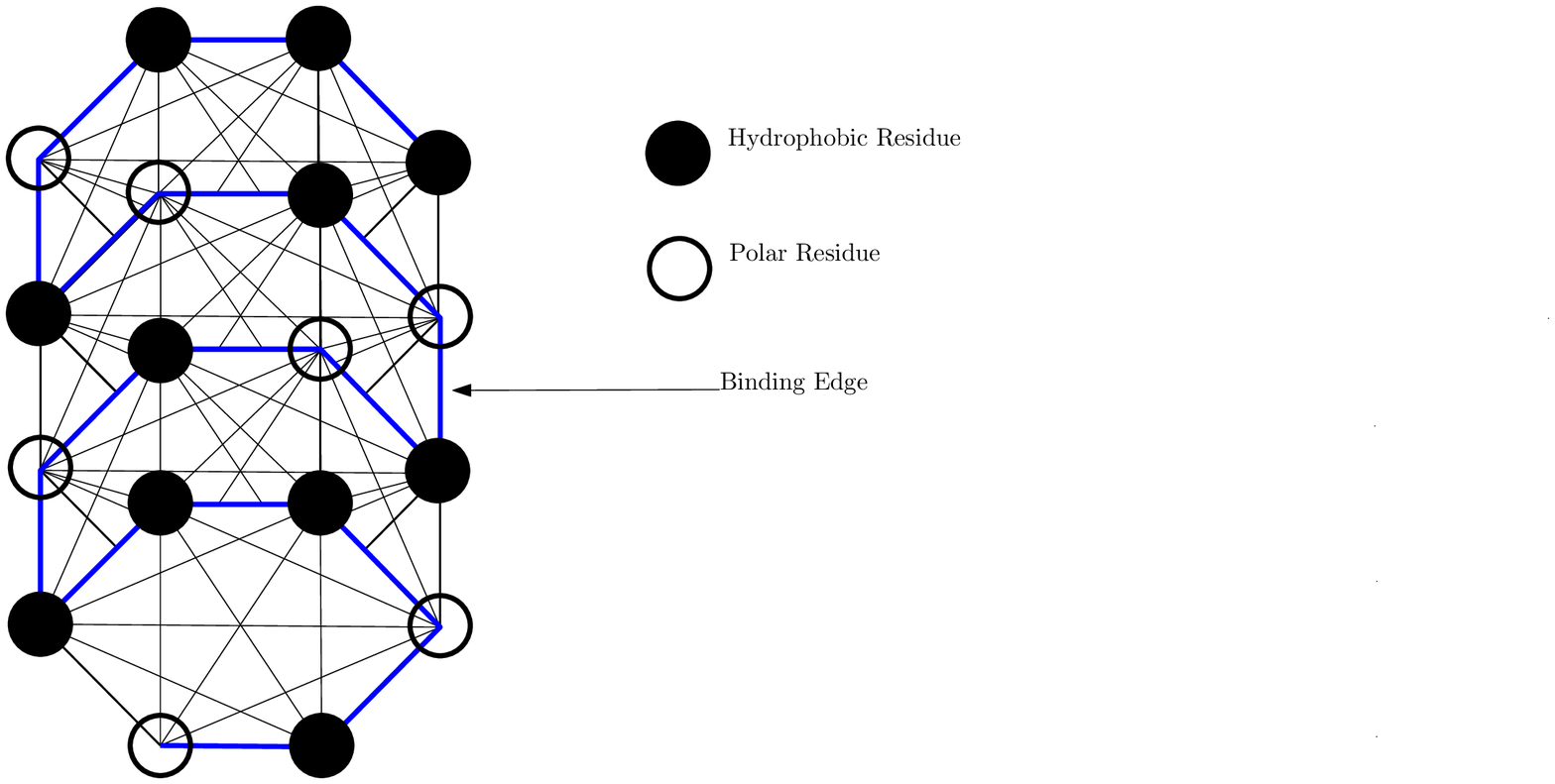,scale=.4}}
\caption{{\normalsize Conformation of PHPHHHPHPHPHPHPHHH on the lattice.}}\label{fig3}
\end{minipage}
\hfill
\begin{minipage}{.4\textwidth}
\centerline
{\epsfig{file=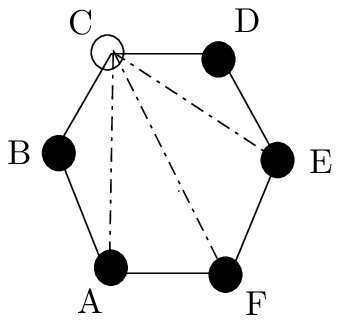,scale=.7}}
\caption{{\normalsize (C,D) and (B,C) are alternating edges; (A,C), (C,F) and (C,E) are loss edges.}}\label{fig4}
\end{minipage}
\end{figure}
In a conformation, a vertex occupied by an H (P) will often be referred to as an H-vertex (a P-vertex). Fig. \ref{fig3} shows an example of a conformation. Edges coloured blue are binding edges and all other edges between residues are non-binding edges. Throughout the paper, the H-vertices are indicated by filled circle and the P-vertices are indicated by blank circles.
\begin{definition}
Given a conformation $ \phi $, an edge $( x,x^{\prime} ) $ of $G$ is called a \textbf{contact edge}, if it is not a binding edge, but there exist $i,j~\in$ $\lbrace 1,\ldots ,t \rbrace $ such that $ f(i) = x, f(j)=x^{\prime},~ and ~ p_i = p_j = H. $ The vertices of the lattice which are not occupied by an H or a P are called \textbf{unused vertices}. A binding edge connecting an H with a P is called an \textbf{alternating edge}. \textbf{Loss edge} is a non-binding edge incident to an H that is not a contact edge (see Fig. \ref{fig4}).
\end{definition}
Now, we define the neighbourhood of an edge in the lattice.
\begin{definition}
Let $ e =(x,y) $ be any edge in $G$. We define the neighbourhood $N(e)$ of $e$ as the intersection of the neighbours of its endpoints $x$ and $y$.
\end{definition}

\section{Our Approaches}\label{OurApproach}
\subsection{Upper Bound}
We will deduce a bound based on a simple counting argument: we will count the number of neighbours of a vertex in the lattice. We start with the following useful lemmas.
\begin{figure}
\centerline
{\epsfig{file=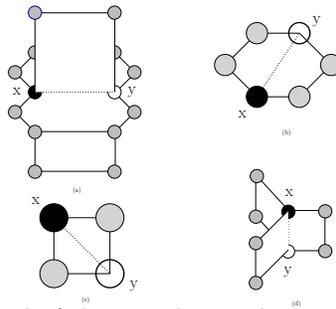,scale=.25}}
\caption{\normalsize(a)12 neighbourhood of the non-diagonal edge $(x,y)$ (b)4 neighbourhood of the diagonal edge $(x,y)$ (c)2 neighbourhood of layer-diagonal edge $(x,y)$ (d)6 neighbourhood of layer non-diagonal edge $(x,y)$.}\label{fig5}
\end{figure}
\begin{lemma}\label{lemma1}
Let $p$ be an HP string and $G=(V,~E)$ is a hexagonal lattice with diagonals. If $p$ has a conformation in $G$, then any H in $p$ can have at most 18 contact edges.
\end{lemma}
  
 \textbf{Proof:} Every vertex in the lattice $G$ has exactly 20 neighbours comprising 3 non-diagonal neighbours, 9 diagonal neighbours in one layer, 4 neighbour from upper layer and 4 neighbour from lower layer (see Fig. \ref{fig1}). In this conformation, every H-vertex has exactly two binding edges. Hence 18 edges remain, which could potentially be contact edges. And hence the result follows. $\qed$ 

\begin{lemma}\label{lemma2}
Let $p$ be an input string for the problem and $ \phi $ be a conformation of $p$. Let $ e $ =$(x,y)$ be a loss edge with respect to $ \phi $. Then there are at most four alternating edges in $N(e)$.
\end{lemma}

\textbf{ Proof:} From Fig. \ref{fig5} if e is a non-diagonal edge, then $N(e)$ contain 12 vertices; if e is a diagonal edge, then $N(e)$ contain 4 vertices; if e is a layer-diagonal edge, then $N(e)$ contain 2 vertices; if e is a layer non-diagonal edge, then $N(e)$ contain 6 vertices. Again, each of $x$ and $y$ can be incident to at most two binding edges. So, there are  at most four binding edges in $N(e)$. It follows immediately that there can be at most four alternating edges adjacent to $e$. $\qed$ 

Now we are ready to present the upper bound.

\begin{lemma}\label{lemma3}
For a given HP string $p$, the the total number of contacts in a conformation $\phi $ is at most $18n - \frac{1}{2} k$, where $k$ is the total number of H-runs and $n$ is the total number of H. 
\end{lemma}

 \textbf{Proof :} From Lemma \ref{lemma1}, we know that the number of contacts is at most $18n$. In a confirmation one loss edge incident to H means that it would lose one contact edge. In what follows we will show that there will be at least $\frac{1}{2} k$ loss edges in $\phi $. Since every H-run is preceded and followed by a total of two alternating edges, it is sufficient to prove that, for each alternating edge in $\phi$ for $p$, we have $\frac{1}{4}$ loss edge on average.\\
From Lemma \ref{lemma2} we know that, for every loss edge there will be at most four alternating edges in its neighbourhood. Alternatively, we can say that, for every four alternating edges there will be at least one loss edge, assuming that the alternating edges are in the neighbourhood of that loss edge. Clearly, if the alternating edges are not within the neighbourhood then the number of loss edges will increase. So, for every alternating edge there will be at least $\frac{1}{4}$ loss edge. There are a total of $2k$ alternating edges. So, the total number of loss edges will be, $\frac{1}{4} \times 2 \times k$ = $\frac{1}{2} k$. Hence, the result follows. $\qed$ 

\subsection{Algorithms and lower bounds}
In this section, we present two novel approximation algorithms for the problem. 

\begin{figure}[h!]
\centerline
{\epsfig{file=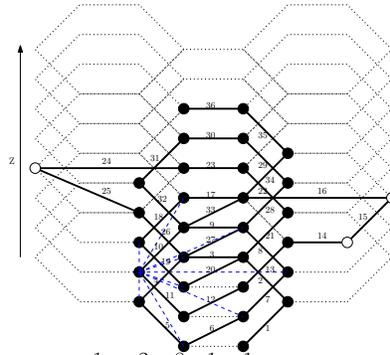,scale=.35}}
\caption{{\normalsize Folding of HP string $H^14P^2H^8P^1H^11$ by Algorithm HelixArrangement. Dotted black line represent the lattice, solid line represent binding edge of protein, blue dashed line shows 9 contacts of a H. Binding edges are numbered sequentially. z indicates the direction of side layers of Upper layer.}}\label{fig55}
\end{figure}

\subsubsection{Algorithm HelixArrangement}

Idea of first algorithm is to arrange all H's of the input string in helix structure. The main difference between conventional helix structure, here we arrange P's of input string outside of the main helix structure. Fig. \ref{fig55} shows the way we arrange H's and P's.

\textbf{Algorithm HelixArrangement} \\
Input: An HP string $p$.
\begin{enumerate}
\item Arrange the H's as follows:
	\begin{enumerate}
	\item Starting from a layer arrange the first six H's in a hexagon. Let, called this base hexagon.
	\item Using the layer diagonal edge climb to upper layer. In this layer arrange next six H's in a hexagon which is parallel to base hexagon. 
	\item repeat step (b) until end of string $p$. The hexagon where the process ended, let called that top hexagon.
	\end{enumerate}
\item Intermediate P-runs are arranged in the outer side of hexagon in a layer(see Fig. \ref{fig55})	
\end{enumerate}

\subsubsection{Approximation ratio for Algorithm HelixArrangement}

Except the H's of base hexagon and top hexagon a H can achieve at least 9 contacts. A H from its layer achieve 3 contacts, from its immediate upper layer 3 contacts and from its immediate lower layer 3 contacts. H's of base hexagon miss the contacts from lower layer and H's of top hexagon miss the contacts from upper layer. So, there is in total 12 H in base hexagon and top hexagon which miss in total $12*3~or~36$ contacts. Note that, it is possible that top hexagon is not filled 6 H's. But it does not change any computation, because there is still 6 H's in top hexagon and lower layer hexagon of top hexagon, which miss 3 contacts. 

Now, if we consider the P's arrangement, we will achieve two contacts for every alternating edge. If there is k alternating edge we will achieve $2k$ contacts.\\
So, for n H's total number of contacts($\mathcal{C}$) can be achieved as follows:
$\mathcal{C} ~ \geq ~ 9n - 36 + 2k$\\
Hence we get the following approximation ratio $A_1$:
\begin{equation}\label{equ0}
A_1=\frac{18n - \frac{1}{2}k}{(9n - 36 + 2k)}
\end{equation}
 From Equation \ref{equ0} it can be seen that for large $n$, $A_1$ tends to reach $\frac{18}{9}~or~2$. So we compute the value of $k$ so that our approximation ratio is at most $2$ as shown below.
 $ \frac{18n - \frac{k}{2}}{(9n-36+2k)} \leq \frac{18}{9}$\\
 $ \Rightarrow 81k \geq 18\times30\times2$  $~~~$ $ \Rightarrow k \geq \frac{40}{3} \approx 13$
 
 So, if the total number of H-runs is greater than $13$, then Algorithm HelixArrangement will achieve an approximation ratio of $2$.

\begin{figure}[h!]
\centerline
{\epsfig{file=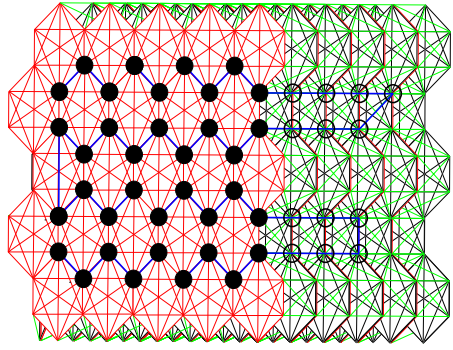,scale=.3}}
\caption{{\normalsize Folding of HP string $H^3P^6H^2P^2H^4P^7H^13P^5H^5P^6H^4P^2H^5$ by Algorithm LayerArrangement only in Upper layer. Z indicates the direction of side layers of Upper layer}}\label{fig6}
\end{figure}

\begin{theorem}\label{theorem0}
For any given HP string, Algorithm HelixArrangement gives a $2$ approximation ratio for $k>13$, where $k$ is the total number of H-runs and $n$ is the total number of H. $\qed$
\end{theorem}

\subsubsection{Algorithm LayerArrangement}
\begin{figure}[h!]
\centerline
{\epsfig{file=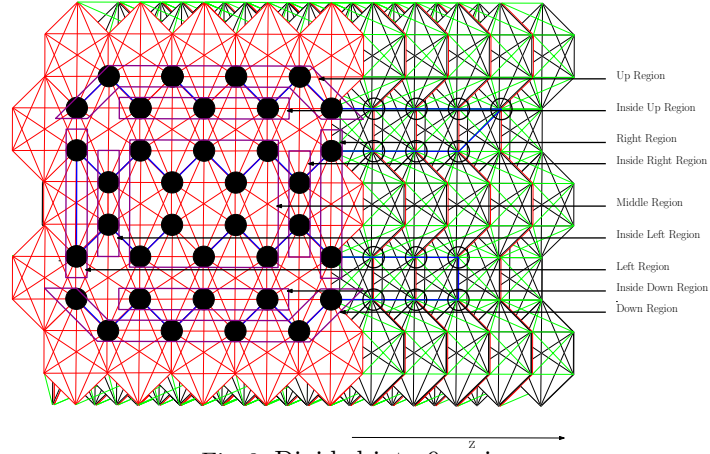,scale=.25}}
\caption{{\normalsize Divided into 9 region}}\label{fig7}
\end{figure}

The idea of second algorithm is to arrange all H's occurring in the input string along the two layers. We arrange the H's in the prefix of the string up to the $\lfloor \frac{n}{2} \rfloor$-th H on the upper layer and arrange the rest of those on the lower layer. In a layer, H-runs are arranged in a spiral manner. Then we arrange the P's between the H's outside these two layers. The arrangements of the P-runs outside the two layers are shown in Fig. \ref{fig6}. Within a layer the arrangement is done in chains (see Fig. \ref{fig6}). The arrangement in the upper (lower) layer can be further divided into nine regions, namely, the left region, the right region, the up region, the down region, the inside-left region, the inside-right region, the inside-up region, the inside-down region and the middle region (see Fig. \ref{fig7}).\\

\textbf{Algorithm LayerArrangement} \\
Input: An HP string $p$.
\begin{enumerate}
\item Set $f$ = $\lfloor \frac{n}{2} \rfloor$.
\item Suppose $F$ denotes the position in $p$ after the $f$-th H. Denote by \textit{pref} $F(p)$ the prefix of $p$ up to position $F$ and by \textit{suff} $F(p)$ the suffix, that starts right after it. Now,
\begin{enumerate}
\item Arrange the H's in \textit{pref} $F(p)$ in the upper layer as follows:
	\begin{enumerate}
	\item Let, $i$ and $j$ are two integers that divide $m_1$ with reminder $0$, such that $|i-j|$ is minimal for all $i$ and $j$. Let, $r=min(i,j)$, which is number of the chains in a layer. Let $s$ = $\lfloor \frac{f}{r} \rfloor$, which is number of residues in a chain. Suppose, $S_1,S_2,S_3...$ denote the position in $p$ after the $s$-th,$2s$-th,$3s$-th$...$ H respectively. So, $S_i(p)$= $p_{S_{i-1}},...,p_{S_{i}-1}$ for $i = 1, 2, 3 ...$. Here $S_0$ is starting position.
	\item Now arrange $S_i(p)$ in chain one by one from top to bottom for $i = 1, 2, 3 ...$.
	\item Intermediate P-runs are arranged in the upper-side layers of the upper layer (see Fig. \ref{fig6})	
	\end{enumerate}
\item Arrange the H's in \textit{suff} $F(p)$ along the lower layer following the same strategy spelled out in Step 2(a); intermediate P-runs are arranged in the lower-side layer of the lower layer (see Fig. \ref{fig6}).
\end{enumerate}
\end{enumerate}

\subsubsection{Approximation ratio for Algorithm LayerArrangement}
Now we focus on deducing an approximation ratio for Algorithm LayerArrangement. Suppose that $m_1= \lfloor \frac{n}{2} \rfloor$. So, according to Algorithm LayerArrangement, the upper (lower) layer will contain $m_1$ ($m_1$ or $m_1+1$) H's.  We consider two cases, namely, where $m_1$ is odd, i.e., $m_1=2x+1$ and $m_1$ is even, i.e., $m_1=2x$, with an integer $x>0$. 

Now, let, $i$ and $j$ are two integers that divide $m_1$ with reminder $0$, such that $|i-j|$ is minimal for all $i$ and $j$. Let, $r=min(i,j)$, which is number of the chains in a layer. Now, let, $s=m_1/r$ which is number of residues in a chain. The chains are arranged spirally in a layer. 

In what follows, we will use $vw$-upper layer ($vw$-lower layer) to denote a particular region of the upper (lower) layer. So, $vw$ could be one of the 9 options, namely, $lR$ (left region), $rR$ (right region), $uR$ (up region), $dR$ (down region), $i_lR$ (inside-left region), $i_rR$ (inside-right region), $i_uR$ (inside-up region), $i_dR$ (inside-down region) and $mR$ (middle region). We also use $\phi_{CA}$ to refer to the conformation given by Algorithm LayerArrangement.

The analysis of this case will be easy to understand with the help of Fig. \ref{fig7}. In $\phi_{CA}$, every vertex in the $lR$-up layer and $rR$-up layer has at least 8 contacts. Every vertex in the $i_lR$-upper layer and the $i_rR$-upper layer has at least 12 contacts. For each of $lR$-upper layer, $rR$-upper layer, $i_lR$-upper layer and the $i_rR$-upper layer, there are $r-2$ such vertices (see Fig. \ref{fig7}). Every vertex in the $uR$-upper layer and the $dR$-upper layer has at least 6 contacts. There are $\frac{s+3}{2}$ such vertices for each of the $uR$-upper layer and the $dR$-upper layer. Every vertex in the $i_uR$-upper layer and the $i_dR$-upper layer has at least 11 contacts. There are $(\frac{s-3}{2})$ such vertices for each of the $i_uR$-upper layer and the $i_dR$-upper layer. So there remain $(rs-2r-2s-4)$ vertices in upper layer which fall to $mR$-upper layer, where every vertex achieved 14 contacts. 

So, the total number of contacts ($\mathcal{C}$ ) of all the vertices of the upper layer can be computed as follows:\\
$ \mathcal{C} \geq 2\times 8\times (r-2) + 2\times 12\times (r-2) + 2\times 6\times \frac{s+3}{2} + 2\times 11\times (\frac{s-3}{2}) + 14\times (2x-2r-2s-4) $\\
$ \Rightarrow \mathcal{C} \geq 16r-32+24r-48+6s+18+11s-33+14sr-28r-28s-56$\\
$ \Rightarrow \mathcal{C} \geq 14sr+12r-11s-151$\\
$ \Rightarrow \mathcal{C} \geq 14m_1+12r-11s-151$ $~~~$ $ \Rightarrow \mathcal{C} \geq 7n+12r-11s-151$\\
Since the upper layer is symmetric to the lower layer, both layer will have the same number of vertices if $n=2m_1$. So all the vertices of the lower layer will also have at least $\mathcal{C}$ contacts. So the total number of contacts will be at least $2\mathcal{C}$ or $14n+24r-22s-302$.

If $n = 2m_1+1$, then let $n_1 = n-1$. This $n_1$ vertices will have at least $14n_1+24r-22s-302$ contacts. The remaining vertex will have at least 2 contacts. So the total number of contacts will be at least $14(n-1)+24r-22s-302+2$ or $14n+24r-22s-314$.So, combining the two cases, we get that the total number of contacts is at least $14n+24r-22s-314$. Now we need to take the alternating edges into our consideration. For every alternating edge we get two extra contacts for the two vertices (each having one). So, for $n$ H's and $k$ alternating edges we get a total of at least $14n+24r-22s-314+2k$ contacts. Hence we get the following approximation ratio $A_2$:
\begin{equation}\label{equ1}
A_2=\frac{18n - \frac{1}{2}k}{(14n+24r-22s-314+2k)}
\end{equation}

From Equation \ref{equ1} it can be seen that for large $n$, $A_2$ tends to reach $\frac{18}{14}$. So we compute the value of $k$ so that our approximation ratio is at most $\frac{18}{14}$ as shown below.\\
 $ \frac{18n - \frac{k}{2}}{(14n+24r-22s-314+2k)} \leq \frac{18}{14}$  \\
 $ \Rightarrow {14}\times {18n - \frac{k}{2}} \leq \frac{18} {(14n+24r-22s-314+2k)}$\\
 $ \Rightarrow 252n - 7k \leq 252n+ 432r - 396s - (314\times 18)+ 36k $\\
 $ \Rightarrow 43k \geq 36(11s-12r) + (314\times 18)$ $~~$ $ \Rightarrow k \geq \frac{36(11s-12r) + (314\times 18)}{43}$\\
 Now, from this case if $11s=12r$, $ k \geq \frac{(314\times 18)}{43} \approx 132$\\
So, if the total number of H-runs is greater than 132, then Algorithm LayerArrangement will achieve an approximation ratio of $\frac{18}{14}$ or $\frac{9}{7}$ for $11s=12r$.
Note that, the value of $k$ is dependent on $n$ and the HP string. We now deduce the expected value of $k$ for a given HP string. This problem can be mapped into the problem of $Integer~ Partitioning$ as defined below. Notably, similar mapping has recently been utilized in \cite{almob}\cite{IWOCA}\cite{BMC} for deriving an expected approximation ratio of another algorithm.

\begin{problem}\label{probsum}
Given an integer $Y$, the problem of Integer Partitioning aims to provide all possible ways of writing $Y$, as a sum of positive integers.
\end{problem} 
Note that the ways that differ only in the order of their summands are considered to be the same partition. A summand in a partition is called a part. Now, if we consider $n$ as the input of Problem~\ref{probsum} (i.e., $Y$) then each length of H-runs can be viewed as parts of the partition. So if we can find the expected number of partitions we could in turn get the expected value of $k$. Kessler and Livingston \cite{Kessler} showed that to get an integer partition of an integer $Y$, expected number of required parts is:
$$\sqrt{\frac{3Y}{2\pi}}\times (\log Y + 2\gamma - 2\log  \sqrt{\frac{\pi}{6}}),$$ where $\gamma$ is the famous Euler's constant. For our problem $Y= n$. If we denote $E[P]$ as the expected number of H-runs then,
$$E[P] = \sqrt{\frac{6}{\pi}}\times \sqrt{n} \times (\frac{1}{2} \log n + \gamma - \log  \sqrt{\frac{\pi}{6}}).$$
Now, as $(\frac{1}{2} \log n + \gamma - \log  \sqrt{\frac{\pi}{6}}) \leq (\sqrt{\frac{2\pi}{3}}\times \frac{1}{2}\log n)$ for $n \geq 5$, we can say that $$E[P] \leq \sqrt{n}\times \log n.$$

So the expected value of $k$ is less than or equal to $\sqrt{n}\times \log n$ which implies that $\sqrt{n}\times \log n \geq 132$ or $n \geq 500$. Now, if $11s > 12r$, lower bound of k increases, as a result expected lower bound of n will increases. On the other side, if $11s < 12r$, expected lower bound of n will decreases. The above findings are summarized in the following theorems. 

\begin{theorem}\label{theorem1}
For any given HP string, Algorithm ChainArrangement gives a $\frac{9}{7}$ approximation ratio for $k>132$, where $k$ is the total number of H-runs and $11s=12r$ where, $n=2rs$ and $n$ is the total number of H. $\qed$
\end{theorem}

\begin{theorem}\label{theorem11}
For any given HP string, Algorithm ChainArrangement is expected to achieve an approximation ratio of $\frac{9}{7}$ for $n\geq500$ and $11s=12r$ where, $n=2rs$ and $n$ is the total number of H. $\qed$
\end{theorem}

\section{Conclusion}
One vertex of SC (Simple Cubic) lattice have 6 neighbour, FCC (Face Centered Cubic) or BCC (Body Centered Cubic) lattice have 14 neighbour. On the other hand, one vertex of hand hexagonal prism lattice with diagonal have 20 neighbour which property leads us to find better approximation ratio. On the other hand this lattice model remove some well known problems of protein folding in SC lattice e.g., parity problem. Considering such properties of this lattice surely tell us that better approximation algorithm could be developed. Also heuristics algorithm can be applied on this lattice, which can surely lead us to better result.

\addcontentsline{toc}{chapter}{Bibliography}
\bibliographystyle{abbrv}
\bibliography{ProteinHPHexa}

%
%
%
%
%
%

\end{document}